\documentclass[11pt]{article}
\usepackage{setspace}
\usepackage{subfigure}
\usepackage{graphicx}
\usepackage{amsmath,amsfonts}
\usepackage{multirow,colortbl,rotating}
\usepackage{natbib}

\renewcommand\today{December 28, 2012}

\DeclareMathOperator{\med}{median}

\newcommand{\data}{{\bf D}}
\newcommand{\param}{\boldsymbol{\theta}}

\newcommand{\mnras}{Monthly Notices of the Royal Astronomical Society}

\begin{document}

\title{A remarkably simple and accurate method for computing the Bayes
  Factor from a Markov chain Monte Carlo Simulation of the Posterior
  Distribution in high dimension}

\author{Martin D. Weinberg\thanks{E-mail: weinberg@astro.umass.edu},
  Ilsang Yoon, and Neal Katz
  \\Department of Astronomy\\University of
  Massachusetts, Amherst, USA}

\date{\today}

\maketitle

\begin{abstract}
  \citet{Weinberg:12b} described a constructive algorithm for
  computing the marginal likelihood, $Z$, from a Markov chain
  simulation of the posterior distribution.  Its key point is: the
  choice of an integration subdomain that eliminates subvolumes with
  poor sampling owing to low tail-values of posterior probability.
  Conversely, this same idea may be used to choose the subdomain that
  \emph{optimizes} the accuracy of $Z$.  Here, we explore using the
  simulated distribution to define a small region of high posterior
  probability, followed by a numerical integration of the sample in
  the selected region using the volume tessellation algorithm
  described in \citet{Weinberg:12b}.  Even more promising is the
  resampling of this small region followed by a naive Monte Carlo
  integration.  The new enhanced algorithm is computationally trivial
  and leads to a dramatic improvement in accuracy.  For example, this
  application of the new algorithm to a four-component mixture with
  random locations in 16 dimensions yields accurate evaluation of $Z$
  with 5\% errors.  This enables Bayes-factor model selection for
  real-world problems that have been infeasible with previous methods.

\end{abstract}

\section{Introduction}
\label{sec:intro}

Bayesian methods hold the promise of selecting general models with
different dimensionality and unrelated structure.  For example,
consider a collection of such models, $\mathcal{M}=\{M_1, M_2, \ldots
M_m\}$, proposed to describe the data $\data$.  Each model $M_j$ is
described by parameter vectors $\param_j\in\mathbb{R}^{d_j}$ where
$d_j = \dim(\param_j)$.  Bayes theorem gives the posterior probability
density for each model:
\begin{equation}
  P(M_j|\data) = \frac{P(M_j) P(\data|M_j)}{P(\data)}
\end{equation}
where $P(M_j)$ is the prior probability of Model $j$, $P(\data)$ is an
unknown normalization and
\begin{equation}
  P(\data|M_j) = \int d\param_j\,P(\param_j|M_j) P(\data|\theta_j,M_j)
\end{equation}
is the marginal likelihood for Model $j$.  The posterior odds of any
two models $j,k\in[1,m]$ is then
\begin{equation}
 \frac{P(M_j|\data)}{P(M_k|\data)} = \left[\frac{P(M_j)}{P(M_k)}\right]
\left[\frac{P(\data|M_j)}{P(\data|M_k)}\right],
\end{equation}
now independent of the normalization $P(\data)$.  The first term on
the right-hand side describes the odds ratio from prior knowledge and
the second term is the \emph{Bayes factor}.  In most cases, the set of
models is given a counting measure; the posterior odds ratio is then
equal to the Bayes factor.  If $P(M_j|\data)\gg P(M_k|\data)$ for all
$k\in[1,m], k\not=j$, Model $j$ \emph{best} explains the data out of
all the proposals in $\mathcal{M}$.

The Bayes factor requires the computation of the marginal likelihood
for each model.  Analytic computation is almost never possible and
direct evaluation by numerical quadrature is almost never feasible for
models of real-world dimensionality and complexity.  This has led to a
variety of approximations based on special properties of the models or
their posterior distributions.  For example, a smooth unimodal
distribution that is well-represented by a multidimensional normal
distribution can be evaluated by Laplace approximation
\citep[e.g.][]{Kass.Raftery:95}.  The dimensionality for nested models
can be effectively lowered as described in \citet{DiCicio.etal:97}.
\citet{Chib.Jeliazkov:2001} describe an efficient approach for models
amenable to block sampling.  A number of astronomical problems of
current interest \citep[e.g.][]{Yoon.etal:2011, Lu.etal:2011,
  Lu.etal:2012} do not fit into these categories and require explicit
methods.

Motivated by these problems, \citet{Weinberg:12b} explored the direct
use of MCMC samples to compute the marginal likelihood and proposed
two algorithms.  In essence, both use the MCMC sample to identify the
important regions of parameter space.  The first algorithm modifies
the harmonic-mean approximation to remove the low-probability tail of
the distribution that dominates the error.  However, if the harmonic
mean integral itself is improper, as it typically is for problems with
weakly informative prior distributions, this algorithm will fail.  The
second algorithm assigns probability to a tree partition of the sample
space and performs the marginal likelihood integral directly.  This
algorithm is consistent for all (proper) posterior distributions.
Additional recent applications have suggested a number of important
extensions to these ideas, which is what we will explore in this
paper.  We will begin in \S\ref{sec:main} with a intuitive motivation
and review of \citet{Weinberg:12b}.

\section{The Main Point}
\label{sec:main}

This contribution focuses on a further extension of the second
algorithm that dramatically improves its feasibility in a variety of
cases.  Let $\Omega$ be the MCMC sample of the desired posterior
probability.  The central point of \citet{Weinberg:12b} is the
following: the marginal likelihood $P(\data|M_j)$ is defined by
\begin{equation}
P(\data|M_j) \int_{\Omega_s} dP(\param_j|\data) = 
\int_{\Omega_s} d\param_j\,P(\param_j|M_j) P(\data|\param_j,M_j)
\label{eq:main}
\end{equation}
where the set $\Omega_s\subset\Omega$ may be chosen to optimize the
numerical evaluations of the integrals in equation (\ref{eq:main}).
Evaluated by Monte Carlo sampling, the integral on the left-hand side
of equation (\ref{eq:main}) is simply the fraction of points in
$\Omega_s$ relative to the number in $\Omega$.  The integral on the
right-hand side of equation (\ref{eq:main}) is performed by quadrature
after the measure is assigned by associating tessellated volume
elements $v({\omega_i})$ in $\mathbb{R}^{d_j}$ to each point or group of
points $\omega_i$ in $\Omega$.  The unity of all subvolumes $v_i
\equiv v(\omega_i)$ in the tessellation is a convex hull in
$\mathbb{R}^{d_j}$.  By construction, the MCMC algorithm provides
samples such that $\int_{\omega_i} d\param_j P(\param_j|\data) \approx
P({\bar\param}|\data) v_i \approx \mbox{constant}$ for some
${\bar\param}\in\omega_i$.  Therefore, relatively small values of
$P({\bar\param})$ will be associated with relatively large values
$v_i$, and, therefore, will contribute most of the variance to the
resulting quadrature on the right-hand side of equation
(\ref{eq:main}).

This motivates seeking subsets $\Omega_s\subset\Omega$ that preserve
the measure defined by the tessellation that minimizes the variance of
equation (\ref{eq:main}).  A particular solution is intuitively
obvious: successively \emph{peel} the subvolumes on the hull until the
volumes $v_i$ are sufficiently small that $P(\param_j|\data)$ varies
slowly across $v_i$ while preserving the condition $|\Omega_s|\gg 1$
so that the error in the integral on the left-hand side of equation
(\ref{eq:main}) remains small\footnote{We denote the cardinality of
  Set $S$ by $|S|$}.  Below, we explore three extensions to this
approach whose goals are optimizing the choice of $\Omega_s$ to
accurately and efficiently evaluate equation (\ref{eq:main}).  In
\S\ref{sec:peeling}, we try the \emph{peeling} approach.  In
\S\ref{sec:peakposterior}, we identify an easy to tessellate volume
near the posterior mode containing the subset $\omega_s$ and
retessellate this volume.  In addition, rather than using the original
MCMC sample, the new subvolume identified from the sample can be
resampled with a new more efficient sampling function, improving the
results.

These proposed extensions do not circumvent the {\it curse} of
dimensionality \citep{Bellman:1957}; we still are required to sample a
high-dimensional space.  However, the approach described above allows
us to choose a domain that results in the most accurate integral with
the smallest number of samples.

\section{Using MCMC for importance sampling choice of 
  the subdomain}

\subsection{Volume peeling}
\label{sec:peeling}

\begin{figure}
  \centering
  \includegraphics[width=0.7\textwidth]{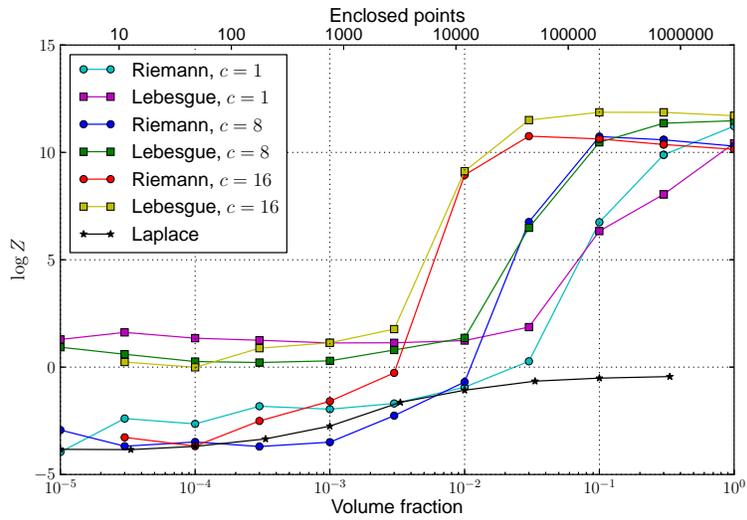}
  \caption{\label{fig:vol12} Values of $\log Z$ obtained with the
    volume-peeling strategy compared with both volume tessellation
    algorithms for a chain of three million states.  The lower axis
    (upper axis) shows the enclosed volume fraction (number of points)
    in the `peeled' subvolume.  The Laplace approximation applied to a
    subset of the entire sample (using the upper axis) is shown for
    comparison.  The true value is $\log Z=0$.}
\end{figure}

We tried two implementations of volume peeling. Both define the
geometric center of mass and slowly decrease the scale of the
self-similar volume in $\mathbb{R}^{d_j}$.  In the first
implementation, We eliminated all subvolumes $v_i$ outside of or
containing the boundary of the scaled self-similar volume.  In the
second, we computed the set $\Omega_s$ contained within the new
boundary and recomputed the tessellation.  The latter is slightly more
accurate than the former but requires a new tessellation.  We adjust
the scale factor to retain a fixed number of points $N$, and,
therefore, limit the error in the left-hand side of equation
(\ref{eq:main}) to a root variance of approximately $1/\sqrt{N}$.

The results of applying the volume peeling strategy to a
Markov-chain-sampled normal distribution in twelve dimensions is
illustrated in Figure \ref{fig:vol12}.  For the tessellation, we use a
k-d tree with a round-robin geometric bisection, sometimes known as
the orthogonal recursive bisection (ORB) tree.  This tree recursively
subdivides the cells into equal subvolumes, one dimension at a
time. The recursion stops when the next division produces cells with
occupation numbers below the target size $c$.  This algorithm prevents
cells with extreme axis ratios but the cells will not have the same
number of counts.  For this test, we adopt a target cell count of
$c=8$.  The figure shows both the Riemann ($\mbox{VTA}_1$) and
Lebesgue ($\mbox{VTA}_2$) variants for estimating the integrals in
equation (\ref{eq:main}), as in described in \citet{Weinberg:12b}.
The Riemann computation uses the median value of the posterior
probability in each cell to compute the contribution to the Riemann
integral:
\begin{equation}
  \int d\param\,P(\param|\data) \approx \sum_j v(\omega_j)
  \med\{P(\param_k|\data)\}
\label{eq:riemann}
\end{equation}
where $\mathbf{\theta}_k\in\omega_j$ and $v(\omega_j)$ is the
hypervolume in the subdomain $j$ (i.e. the cell).  The Lebesgue
computation assigns the hypervolume by probability according to the
monotone function $f(P)$ as follows:
\begin{equation}
v(\omega_j; P) \equiv \int_{P>P_k; k\in\omega_j} d\param \approx 
\frac{v(\omega_j)}{|\omega_j|}\sum_{k\in\omega_j}
\begin{cases}
  1 & \mbox{if}\, P \geq P_{k+1} \\
  \frac{f(P) - f(P_k)}{f(P_{k+1}) - f(P_k)} & \mbox{else if}\, P_{k+1} > P > P_k, \\
  0 & \mbox{otherwise}
\end{cases}
\label{eq:lebesgue}
\end{equation}
where we assume that $P_l > P_k$ if $l > k$.  For the computations
here, we choose $f(x) = x$.  The Lebesgue integral is then
approximated as
\[
\int d\param\,P(\param|\data) \approx \sum_k \frac12\left(P_{k+1} -
  P_{k}\right)\left(\sum_j v(\omega_{k+1}) + \sum_j v(\omega_{k})\right).
\]
The Riemann and Lebesgue constructions differ, even in the limit
$c=1$.

Figure \ref{fig:vol12} demonstrates that the choice of an appropriate
subdomain results in acceptable approximations to the marginal
likelihood.  In all cases, the sparsely sampled tails of the
distribution coupled with regions of empty volume bias the evaluation
of $Z$ upward.  As the volume is peeled from the outside, the offset
from the true value $Z=1$ ($\log Z=0$) decreases dramatically, as
expected.  All evaluations are for a fixed chain size. Therefore,
larger values of $c$ imply lower spatial resolution.  In addition, as
$c$ increases and the cell width approaches the characteristic scale
of the posterior distribution, the variance of $P$ in the cell
increases and the approximations in equations (\ref{eq:riemann}) and
(\ref{eq:lebesgue}) become inaccurate.  On the other hand, a small
volume fraction constrains the sampled region to the peak of the
distribution.  Near the peak, the variance of $P$ decreases and
accuracy is recovered.  This explains the shift to smaller volume
fraction required to obtain accuracy with higher $c$ in Figure
\ref{fig:vol12}.  Both variants are biased in the limit of small
volume fractions; the Lebesgue (Riemann) method is biased high (low).
The Lebesgue (Riemann) variant with $c\ge8$ ($c=1$) has the smallest
bias. The Lebesgue variant has the lowest bias overall.

The underestimation of $Z$ by the Riemann algorithm makes intuitive
sense; the distribution of probability values in each cell will be
exponentially skewed toward higher values.  Using the median value as
the representative value for the cell will tend to underestimate the
cell's true contribution.  The overestimation of $Z$ by the Lebesgue
owes to the linear assignment of probability to volume in the measure
function (our choice of $f(x)$ in eq. \ref{eq:lebesgue}); the true
value of $f(x)$ is likely to be some convex-up function.

We show the Laplace approximation for comparison.  We use the same
number of enclosed points implied by the volume fraction for the
Riemann and Lebesgue variants but randomly sampled from the entire
chain with no volume restriction.  As expected, the Laplace
approximation converges to the true value for a large sample from a
multivariate normal distribution. Similar results obtain for lower and
higher dimensionality.  We have checked this up to $d=16$.

Applying the Lebesgue variant of the volume tessellation algorithm
(VTA) to an appropriately `peeled' volume for a unimodal distribution
provides a usefully accurate evaluation of the marginal likelihood.
Now imagine a bimodal distribution with two widely separated modes.
The volume peeling strategy will fail miserably: the central
fractional volume will tend to sit in the poorly sampled desert
between the two modes.  An adaptive approach is needed.

\subsection{Identification of the posterior mode}
\label{sec:peakposterior}

The previous family of subdomain selections relies purely on the shape
of the enclosing volume.  In general, this algorithm will not center
the subdomain on the shallowest part of the posterior distribution.
For a worst-case counterexample, consider two equally shaped but
widely separated spherical modes in parameter space.  The algorithm in
\S\ref{sec:peeling} will eliminate the peaks and retain the tails of
both modes.  This leads to a worse estimate than the original estimate
based on the full posterior sample.

This counterexample suggests the following alternative approach:
\begin{enumerate}
\item Set the center to the location of the parameter point $\param_k$
  with the maximum value of the posterior probability:
  $P(\param_k|\data)>P(\param_j|\data)$ for all $j\not=k,
  j\in[1,\ldots,N]$

\item Compute the shape of the hyperrectangle initially from the
  parameter ranges of the entire sample: $\sigma_{0r} =
  \max\{\theta_{1r},\ldots,\theta_{Nr}\} -
  \min\{\theta_{1r},\ldots,\theta_{Nr}\}$.

\item Let $q$ count the number of iterations and set $q=0$ to start.

\item Compute the distances of the sample from $\param_k$:
  $d_j^2 = \sum_r(\theta_{jr} - \theta_{kr})^2/\sigma_{qr}^2$
  \label{itm:dist}

\item Let ${\bar d}$ be the $M^{\mbox{th}}$ distance in the sorted
  list of distances.  Choose $M$ large enough to achieve a
  sufficiently small variance in ${\bar d}$ (e.g. $M=10^3$).  The
  enclosing hyperrectangle now has the coordinates $\theta_{min,r} =
  \theta_{kr} - \sigma_{0r}{\bar d}$ and $\theta_{max,r} = \theta_{kr}
  + \sigma_{0r}{\bar d}$.
  \label{itm:bard}
  
\item Increment $q$ and recompute the shape of the hyperrectangle from
  the variance of the entire sample, $\sigma_{qr}^2 = \sum_j
  (\theta_{jr} - \theta_{kr})^2$ for $\param_j \in
  [\param_{min}, \param_{max}]$.  Repeat Steps
  \ref{itm:dist}--\ref{itm:bard}.
  \label{itm:var}

\item Steps \ref{itm:dist}--\ref{itm:var} may be iterated until
  converged, if desired.

\item Finally, tessellate the volume defined by these $M$ points and
  compute the right-hand side of equation (\ref{eq:main})
  \label{itm:compute}

\end{enumerate}

This algorithm significantly improves the marginal likelihood
computations, resulting in errors in the log of the marginal
likelihood of several tens of percent, i.e. $|\delta\log
P(\data|M_j)|<0.2$. However, the bias in these estimates appears to
decrease slowly with sample size.  This bias appears to result from
large changes in $P(\param_j|\data)$ across the subvolumes.  To test
this speculation, we replaced Step \ref{itm:compute} in the algorithm
listed in \S\ref{sec:peakposterior} by a uniform resampling of the
hyperrectangle.  A uniform sampling over the volume prevents the
volume of the cells growing as the probability value decreases and
results in a strong suppression of the bias.  A sampling function with
less variance than uniform over the sample volume might lead to better
accuracy while still suppressing the bias, but we have not
investigated this possibility.

\subsection{Test problems}

\begin{figure}[thb]
  \subfigure[separated]{
    \includegraphics[width=0.49\textwidth]{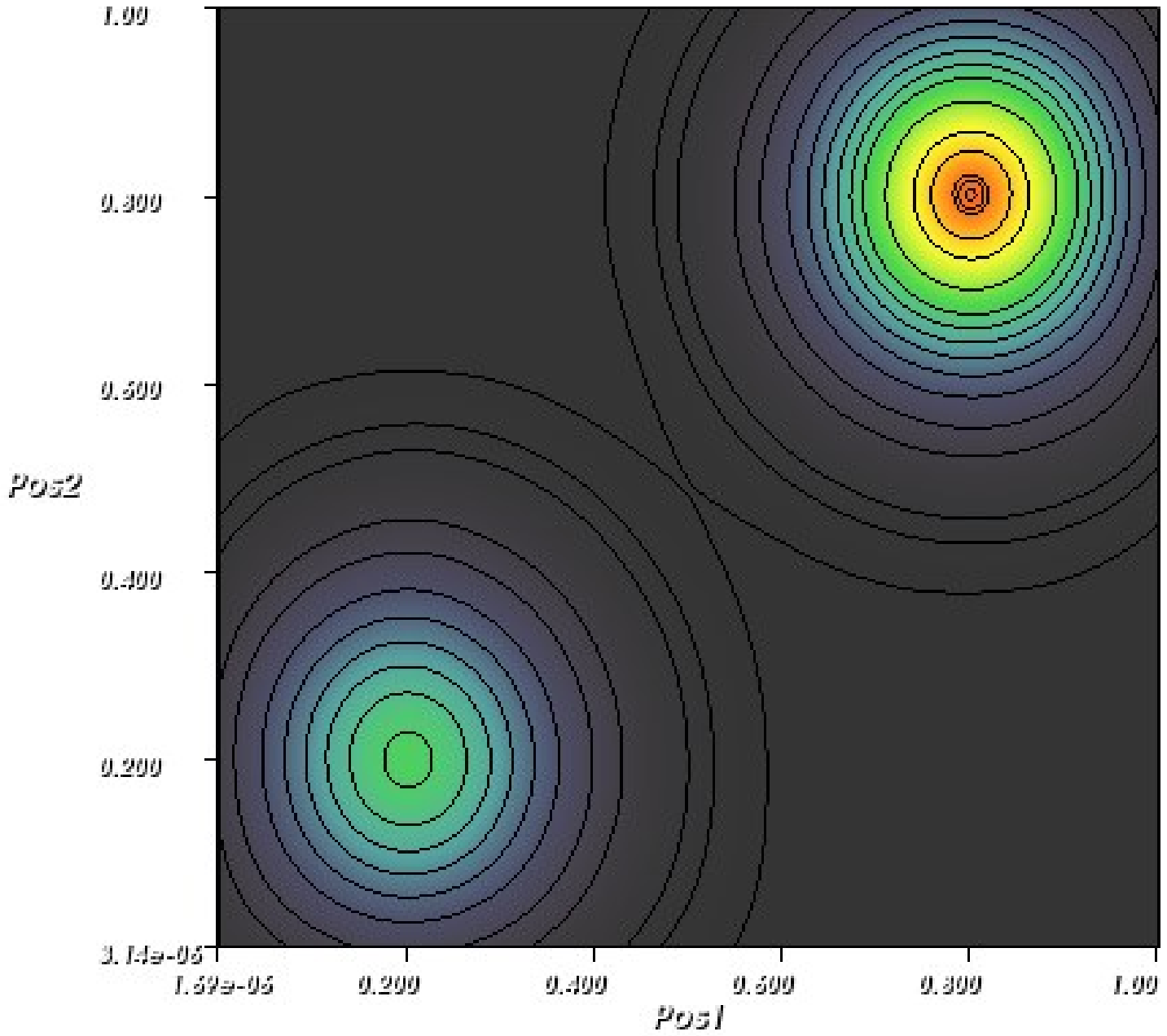}
  }
  \subfigure[overlapped]{
    \includegraphics[width=0.49\textwidth]{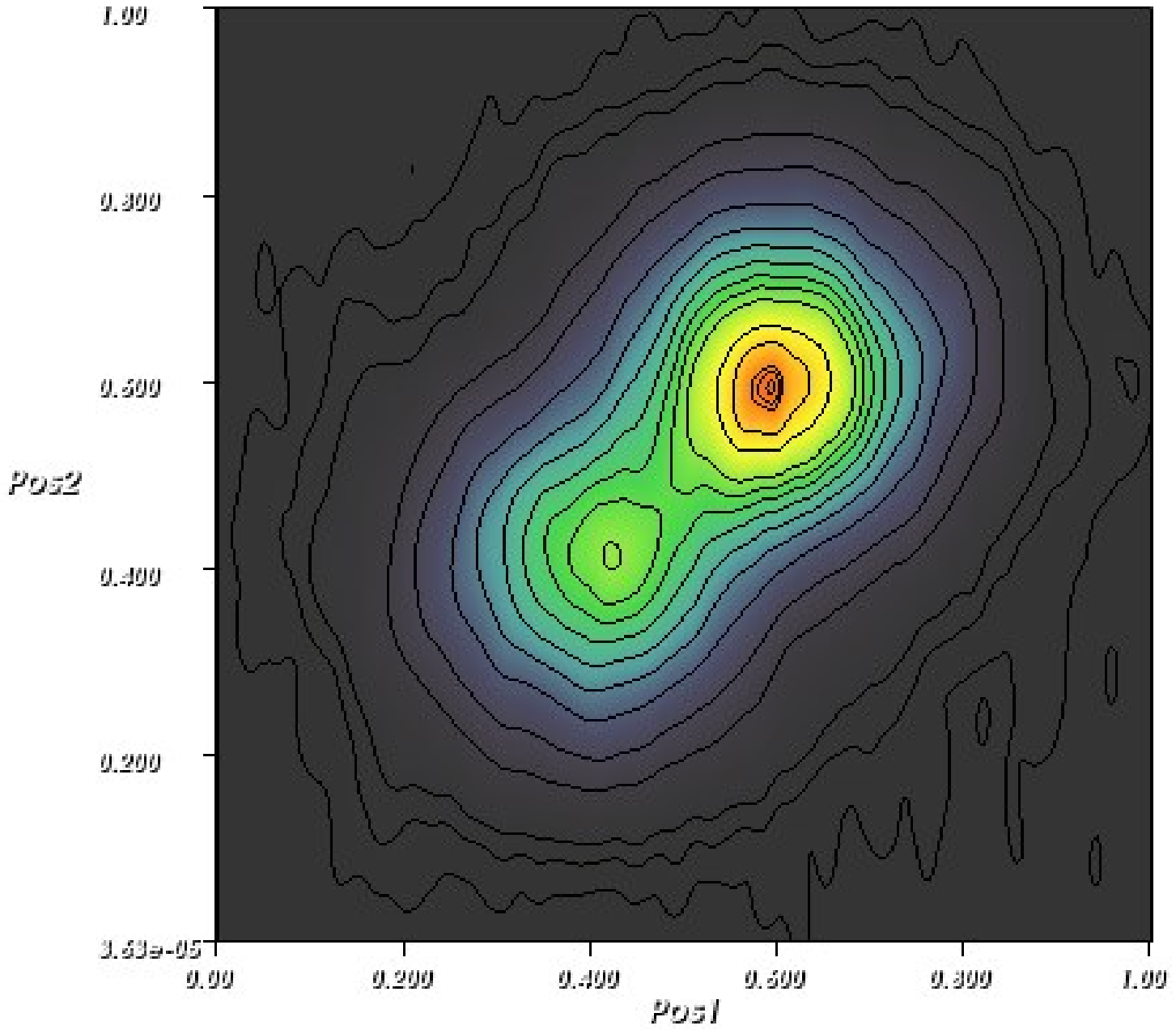}
  }
  \caption{Marginalization of two-component Gaussian distribution in
    8 dimensions using $10^5$ MCMC samples for Test Distribution 2
    (left) and Test Distribution (3) (right).  The contour levels are
    0.001, 0.005, 0.01, 0.05, 0.1, 0.2, 0.3, 0.4, 0.5, 0.6, 0.7, 0.8,
    0.9, 0.95, 0.99, 0.995, 0.999. \label{fig:twocomp_proj}}
\end{figure}

\begin{figure}[thb]
  \includegraphics[width=0.8\textwidth]{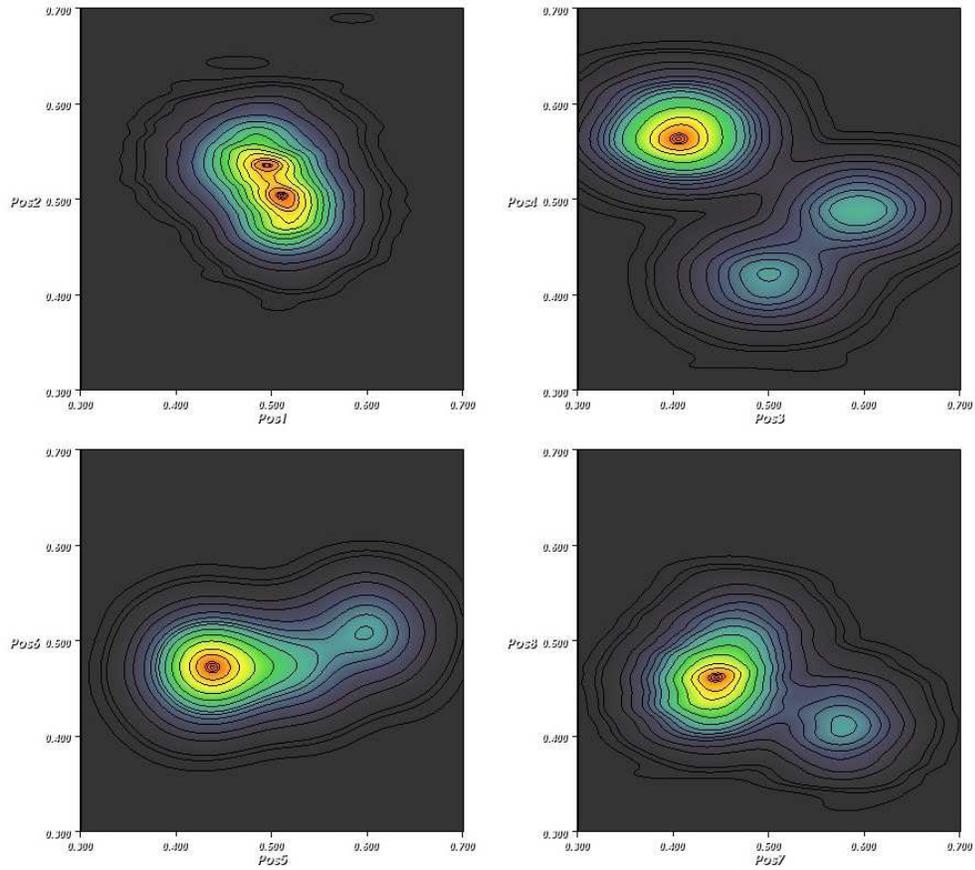}
  \caption{\label{fig:blobs} Marginalized distribution of the
    four-component Gaussian distribution with randomly chosen centers
    in 8 dimensions using $10^5$ MCMC samples for Test Distribution 4.
    The four panels describe different projections as indicated by the
    axis labels.  Contour levels are as described in
    Fig. \protect{\ref{fig:twocomp_proj}}.}
\end{figure}

Here, the original and new algorithms are applied to the following
four test distributions constructed from one or more normal components
with a variance $\sigma^2=0.003$ and centered in the unit hypercube of
dimensionality $d$.  The centers for the test distributions are as
follows:
\begin{enumerate}
\item A \emph{single} distribution with center ${\bar x}=(0.5,\ldots,0.5)$.
\item Two \emph{separated} distributions with centers ${\bar x}_1 =
  (0.2, 0.2, 0.5,\ldots,0.5)$ and ${\bar x}_2 = (0.8, 0.8,
  0.5,\ldots,\allowbreak 0.5)$ with corresponding weights $w_1=0.6,
  w_2=0.4$.  This distribution simulates two widely separated modes.
  Figure \ref{fig:twocomp_proj} (left) shows the distribution
  marginalized in all but the first two dimensions for $d=8$ and for a
  MCMC-generated sample of $10^5$ points.
\item Two \emph{overlapping} distributions with centers ${\bar x}_1 =
  (0.4, 0.4, 0.5,\ldots,0.5)$ and ${\bar x}_2 = (0.6, 0.6, 0.5,
  \ldots,\allowbreak 0.5)$ with corresponding weights $w_1=0.6$,
  $w_2=0.4$.  This distribution a distribution with two local maxima
  on a common pedestal.  Figure \ref{fig:twocomp_proj} (right) shows
  shows the distribution marginalized in all but the first two
  dimensions for $d=8$ and $10^5$ MCMC points.
\item Four \emph{randomly-oriented} distributions with their centers
  uniformly selected from the hypercube $[0.5-2\sigma, 0.5+2\sigma]^d$
  at random with Dirichlet distributed weights and shape parameter
  $\alpha=1$.  This produces an asymmetric distribution with multiple
  maxima connected by ``necks'' of varying amplitude and emulates
  features of posterior distributions from parametric models found in
  practice \citep[e.g.][]{Lu.etal:2012}.  Figure \ref{fig:blobs} shows
  the distribution in pairs of marginal variables for $d=8$.
\end{enumerate}

\begin{sidewaystable}
  \centering
  \caption{Error in marginal likelihood values: original algorithm}
  \begin{minipage}{0.7\textwidth}
    \begin{tabular}{|r|r|r|r|r|r|r|r|r|}
      \hline
      \multicolumn{2}{|>{Model}c|}{}&\multicolumn{7}{|>{Error in $\log
          Z$}c|}{}\\
      \cline{3-9}
      \multicolumn{2}{|>{}c|}{}&\multicolumn{3}{c|}{100\% Volume, $c=8$}&%
      \multicolumn{2}{c|}{0.1\% volume, $c=8$}&\multicolumn{2}{c|}{0.1\%
        volume, $c=16$}\\
      \cline{3-9}
      \hline
      Type & d  
      &$\hbox{VTA}_1$\footnote{Riemann evaluation}
      &$\hbox{VTA}_2$\footnote{Lebesgue evaluation}
      &Laplace&$\hbox{VTA}_1$&$\hbox{VTA}_2$
      &$\hbox{VTA}_1$&$\hbox{VTA}_2$ \\
      \hline
      \multirow{3}{*}{1. Single}&4
      & $2.68$ & $0.55$ & $0.00$
      & $-3.11$ & $-0.07$ 
      & $-3.21$ & $-0.08$ \\
      &8
      & $6.38$ & $5.03$ & $-0.08$
      & $-2.56$ & $0.06$ 
      & $-2.70$ & $0.01$ \\
      &12 
      & $9.33$ & $10.24$ & $-0.29$ 
      & $-3.49$ & $0.29$
      & $-1.59$ & $1.13$ \\
      &16 
      & $14.10$ & $11.58$ & $0.66$
      & $13.22$ & $15.59$
      & $12.75$ & $14.96$ \\
      \hline 
      \multirow{3}{*}{2. Separated}&4
      &$10.46$ & $9.10$ & $9.33$
      & $\infty$ & $\infty$ & $\infty$ & $\infty$ \\
      &8 
      &$10.78$ & $9.73$ & $8.40$
      & $\infty$ & $\infty$ & $\infty$ & $\infty$ \\
      &12 
      & $11.07$ & $8.70$ & $1.12$
      & $\infty$ & $\infty$ & $\infty$ & $\infty$ \\
      &16 & $13.00$ & $10.51$ & $3.49$
      & $\infty$ & $\infty$ & $\infty$ & $\infty$ \\
      \hline
      \multirow{3}{*}{3. Overlapped}&4
      &$3.08$ & $0.76$  & $0.49$
      & $-4.60$ & $-0.03$ & $-3.55$ & $-0.13$ \\
      & 8
      & $6.51$ & $5.35$ & $0.12$
      & $-3.35$ & $0.29$ & $-3.55$ & $0.16$ \\
      &12 
      & $9.44$ & $10.11$ & $-0.71$
      & $-3.61$ & $0.40$ & $-3.78$ & $0.20$ \\
      &16 & $12.65$ & $13.87$ & $-0.64$
      & $5.36$ & $6.38$ & $14.25$ & $15.76$ \\
      \hline
      \multirow{3}{*}{4. Random}
      &4 
      &$-2.50$ & $-0.01$ & $4.76$
      &$-2.51$ & $ 0.02$ & $-4.26$ & $0.12$ \\
      &8 
      & $-1.37$ & $0.67$ & $4.41$
      & $-1.13$ & $0.92$ & $-1.92$ & $0.99$ \\
      &12
      & $1.37$ & $3.81$ & $4.40$ 
      & $7.99$ & $9.79$ & $-1.24$ & $2.42$ \\
      &16
      &$-15.76$ & $16.57$ & $4.67$
      & $14.91$ & $16.20$ & $14.33$ & $17.06$ \\
      \hline
    \end{tabular}
  \end{minipage}
  \label{tab:logZorig}
  \end{sidewaystable}

\begin{sidewaystable}
  \centering
  \caption{Error in marginal likelihood values: important region}
  \begin{tabular}{|r|r|r|r|r|r|r|r|}
    \hline
    \multicolumn{2}{|>{Model}c|}{}&\multicolumn{6}{|>{Error in $\log Z$}c|}{}\\
    \cline{3-8}
    \multicolumn{2}{|>{}c|}{}&
    \multicolumn{3}{c|}{Subregion}&\multicolumn{3}{c|}{Resampled}\\
    \cline{3-8}
    \hline
    Type & d
    &$\hbox{VTA}_1$&$\hbox{VTA}_2$&Mean
    &$\hbox{VTA}_1$&$\hbox{VTA}_2$&Mean\\
    \hline
    \multirow{3}{*}{1. Single}&4
    & $-0.090$ & $-0.093$ &
    $-0.0092$  & $-0.014$ & $-0.023$ &
    $-0.016$ \\
    &8
    & $-0.667$ & $-1.097$ & $-0.676$
    & $-0.081$ & $-0.121$ & $-0.0954$ \\
    &12 
    & $-0.746$ & $-1.178$ & $-0.795$
    &$0.056$&$-0.021$&$-0.018$ \\
    &16 
    & $-0.43$ & $-1.08$ & $-0.48$
    & $-0.13$ & $-0.12$ & $-0.09$ \\
    \hline
    \multirow{3}{*}{2. Separated}&4
    &$-0.472$ &  $-0.484$& $-0.475$
    &$-0.040$ & $-0.032$ &
    $-0.036$ \\
    &8 
    &$-0.710$ & $-1.089$ & $-0.703$
    &$-0.115$ & $-0.163$ &
    $-0.140$ \\
    &12 
    & $-0.667$ & $-1.41$ & $-0.717$
    & $-0.121$ & $-0.200$ &
    $-0.139$ \\
    &16
    &$-0.702$  & $-0.702$ & $-0.571$
    &$0.027$&$-0.161$&$-0.141$ \\
    \hline
    \multirow{3}{*}{3. Overlapped}&4
    &$-0.389$ &  $-0.389$& $-0.391$
    &$0.108$ & $0.102$ &
    $0.105$ \\
    & 8
    & $-0.002$ & $-0.076$ & $-0.020$
    & $0.075$ & $0.016$ &
    $0.053$ \\
    &12 
    &$-0.445$1 & $-0.548$ & $-0.515 $
    &$0.070$&$-0.021$&$0.0178$ \\
    &16
    &$-0.455$  & $-0.846$ & $-0.520$
    &$0.200$&$-0.031$&$0.048$ \\
    \hline
    \multirow{3}{*}{4. Random}
    &4 
    &$-0.460$&$-0.466$&$-0.463$
    &$0.099$&$0.044$&$0.044$ \\
    &8 
    &$-0.456$&$-0.502$&$-0.499$
    &$0.021$&$0.044$&$0.044$ \\
    &12
    & $-0.595$&$-0.704$&$-0.628$
    &$0.099$&$-0.081$&$-0.080$ \\
    &16
    &$-0.595$ & $-0.740$ & $-0.628$
    &$0.129$&$0.069$&$0.070$ \\
    \hline
  \end{tabular}
  \label{tab:logZnew}
\end{sidewaystable}

Table \ref{tab:logZorig} summarizes the marginal likelihood evaluation
using the original VTA algorithm with the ORB tree as described in
\citet{Weinberg:12b} with volume trimming.  In all cases, the input
chain has $2\times10^5$ states.  The first group of columns lists the
model, from the list above, and dimensionality.  The second group of
columns lists $\log Z$ using the entire MCMC sample from $\Omega$ for
the Riemann ($\mbox{VTA}_1$), Lebesgue ($\mbox{VTA}_2$) evaluations
using volume tessellation and the Laplace method.  For all but the
lowest dimensionality, the resulting value of $Z$ is biased upward as
for reasons discussed previously.  This is designed for comparison
with the third and fourth groups of columns computed for the fraction
0.001 of the original tessellated volume and $c=8$ and $c=16$,
respectively.

Most notably, the model with two \emph{separated} normal distributions
has no samples in the subvolume between the two modes.  For the other
three cases, the trimmed volume leads to improved accuracy.  In most
cases, the Lebesgue evaluation ($\hbox{VTA}_2$) significantly
outperforms the Riemann evaluation ($\hbox{VTA}_2$) owing to the extra
information about the distribution of posterior probability values in
each cell.  In all cases, the posterior distribution is undersampled
for $d=12$ and $d=16$ leading to poor results, with some errors
greater than 14 in the log!  Even when the volume trimming is centered
on the posterior mode as it is for Test Distribution 1, the sparse
sampling of the mode for $d=12, 16$ and fixed sample size yields large
inaccuracies.

Contrast the results from Table \ref{tab:logZorig} with those from
Table \ref{tab:logZnew}, which illustrates the effect of identifying a
subvolume around the peak posterior value as described in
\S\ref{sec:peakposterior} with $M=1000$.  This choice implies a
relative sampling error of $1/\sqrt{1000} = 0.03$.  The posterior
samples are identical in both cases.  In Table \ref{tab:logZnew}, the
second group of columns (identified by `Subregion') lists the errors
in $\log Z$ evaluated by retessellating the $M$ \emph{important}
points and the third group of columns (identified by `Resampled') lists
the errors in $\log Z$ evaluated by resampling the important region
uniformly in each dimension using $3\times10^5$ points.  The column
labeled `Mean' denotes the average value of the points in the
important region multiplied by the volume of the important region.
This is equivalent to the naive Monte Carlo evaluation of $Z$ for the
resampled case, which we show for the retessellated case for
comparison.

In nearly all cases, the new algorithm outperforms the original one,
and significantly so for high values of $d$.  The accuracy of the
resampled case is remarkable: the computed values are within 25\% of
the exact value in all cases and much smaller in most cases.  No
significant differences are found between the Riemann, Lebesgue, and
naive Monte Carlo evaluations, owing to the slowing varying values of
the posterior probability $P(\data|\param_j,{\cal M}_j)$ across the
important region.  For the subregion evaluation based on the MCMC
posterior sample, the results are clearly worse with $\cal{O}(1)$
errors, but are still remarkably better for high dimensionality.  For
lower dimensionality, the original algorithm with volume trimmer out
performs the new algorithm where the subregion contains enough points
to make an evaluation possible.

In summary, using the MCMC posterior sample to identify an
\emph{important region} around a dominant mode and resampling this
region to evaluate the integral on the right-hand side of equation
(\ref{eq:main}) yields accurate results with a modest number of
evaluations for dimensionality $d\le16$.  Moreover, using the MCMC
chain to evaluate the integral on the left-hand side of equation
(\ref{eq:main}) and the naive Monte Carlo evaluation on the right-hand
side yields an accurate result without the more elaborate volume
tessellation.  These results obtain even for a random distribution of
four connected modes (see Fig. \ref{fig:blobs}).

\section{Summary and Discussion}

Markov chain Monte Carlo sampling of posterior distributions is a
common tool in Bayesian inference, particularly for parameter
estimation.  These samples, often obtained for suites of competing
complex models often require a substantial investment in computational
resources.  This motivates reusing the sample whenever possible.  For
example, it would be wonderful to exploit Bayesian model selection
techniques, Bayes factors in particular, without a new costly
computational campaign or resorting to often inaccurate
approximations.

Motivated by this desire, \citet{Weinberg:12b} presented two
algorithms for computing the Bayes normalization or \emph{marginal
  likelihood} value using MCMC-sampled posterior distributions.  The
main point of that paper was that the numerical evaluation of the
integrals in equation (\ref{eq:main}) that defines the normalization
$P(\data|\cal{M})$ can be performed for a subdomain
$\Omega_s\subset\Omega$ and equation (\ref{eq:main}) still holds.  Our
current paper suggests using an appropriately defined subset to
eliminate volume elements with large errors.  The main point is that a
subsample with $|\Omega_s|\ll|\Omega|$ but centered around a posterior
mode renders the integrals in equation (\ref{eq:main}) as accurately
as possible.

Moreover, and much to our amazement, by resampling the small subdomain
$\Omega_s$, accurate values for the normalization $P(\data|\cal{M})$
are obtained by estimating the fraction of the sample in $\Omega_s$
using the original MCMC-generated posterior sample to evaluate the
integral on the left-hand side of equation (\ref{eq:main}) and using a
uniform resampling of volume defined by $\Omega_s$ followed by a naive
Monte Carlo integration estimate of the right-hand side of equation
(\ref{eq:main}).  This may be done without the elaborate tessellation
algorithm defined in \citet{Weinberg:12b}!

The sample sizes required are still bound by the curse of
dimensionality.  In particular, the evaluation of the left-hand side
is a counting process with $M$ points and the Poisson error is
proportional to $1/\sqrt{M}$.  An accurate evaluation of the
right-hand side requires that the posterior probability be as uniform
as possible in the subvolume.  This requires an ever larger posterior
sample as the dimensionality $d$ increases.  Fortunately, this
condition can be easily diagnosed as part of the computation.

This work suggests a number of future improvements.  For example, the
error analysis could be automated by using a stopping criterion for
the initial posterior sample selection that enforces a predefined
number of samples $M$ in a volume with
$\max\{P(\param|\data)\}/\min\{P(\param|\data)\} < L$ with $L$ chosen
such that the integral will converge quickly by cubature once the
``core'' of the posterior mode is reached.  Other possible avenues
include using sampling by quasi-random numbers or importance sampling
based on the covariance matrix of the samples in the subdomain.  A
paper currently in preparation, will describe the application of this
algorithm to marginal likelihood values used to classify astronomical
images \citep{Yoon.etal:2011,Yoon.etal:2013a,Yoon.etal:2013b}.

\section*{Acknowledgments}

This work was supported in part by the NASA AISR Program through award
NNG06GF25G and NSF awards 0611948, 1009652, 1109354, and the
University of Massachusetts/Amherst.

\bibliographystyle{apalike}
\bibliography{mnemonic,master,suppl}

\label{lastpage}

\end{document}